# A FUZZY COMMITMENT SCHEME

**Alawi A. Al-saggaf**
Computer Science and Engineering College
Al-Ahgaff University – Hadhramout
Republic of Yemen
alwiduh@yahoo.com
**Acharya H. S.**
Symbiosis Institute of Computer Studies and Research
Symbiosis International University – Pune-India
haridas.acharya@symbiosiscomputers.com

**ABSTRACT**
*This paper attempt has been made to explain a fuzzy commitment scheme. In the conventional Commitment schemes, both committed string m and valid opening key are required to enable the sender to prove the commitment. However there could be many instances where the transmission involves noise or minor errors arising purely because of the factors over which neither the sender nor the receiver have any control.*
*The fuzzy commitment scheme presented in this paper is to accept the opening key that is close to the original one in suitable distance metric, but not necessarily identical. The concept itself is illustrated with the help of simple situation.*
**KEY WORDS**
Cryptography, Error Correcting Codes, Fuzzy logic and Commitment scheme.

## 1. Introduction

The notion of Commitment scheme is at the heart of most the constructions of modern Cryptography protocols. Protocols are essentially a set of rules associated with a process or a scheme defining the process. Commitment schemes are the processes in which the interests of the parties involved in a process are safeguarded and the process itself is made as fair as possible. Commitment protocols were first introduced by Blum [1] in 1982; many more Commitment Schemes were later developed with improved features [5, 6, 7, 8, 12, 13]. Moreover in the conventional Commitment schemes, opening key are required to enable the sender to prove the commitment. However there could be many instances where the transmission involves noise or minor errors arising purely because of the factors over which neither the sender nor the receiver have any control.
Our aim in this paper to describe commitment schemes, which use algorithms to counter possible uncertainness. Uncertainty leads to introduction of fuzzy sets and fuzzy logic[2] in to the protocol itself.

*Fuzzy commitment* scheme was first introduced by Juels and Martin [3], *fuzziness* also introduced later in [4,14,15] for generating cryptographic keys. They add new property called "*fuzziness*" in the open phase to allow, acceptance of the commitment using corrupted opening key that is close to the original one in appropriate metric or distance. In this paper we have attempted a more formal and mathematical definition of fuzzy commitment schemes. An overview of commitment schemes and description of related work is also incorporated. A brief introduction of error correcting codes, with real life situation to illustrate is attempted.

## 2. Crisp Commitment Schemes

In a conventional commitment scheme, one party, whom we denote the sender namely Alice, aim to entrust a concealed message m to the second party namely Bob. Intuitively a commitment scheme can be seen as the digital equivalent of a sealed envelope. If Alice wants to commit to some message m she just puts it into the sealed envelope, so that whenever Alice wants to reveal the message to Bob, she opens the envelope. Clearly, such a mechanism can be useful only if it meets some basic requirements. First of all the digital envelope should hide the message from: Bob should be able to learn m from the commitment (this is often referred in the literature as the hiding property). Second, the digital envelope should be binding, meaning with this that Alice can not change her mind about m, and by checking the opening of the commitment one can verify that the obtained value is actually the one Alice had in mind originally (this is often referred to as the binding property).

**Definition 1:** *A Commitment scheme* is a tuple *{P, E,M }* Where $M = \{0,1\}^n$ is a message space, *P* is a set of individuals, generally with three elements A as the committing party, B as the party to which Commitment is made and TC as the trusted party, $E = \{(t_i, e_i)\}$ are called the events occurring at times

$t_i$, i = 1,2,3, as per algorithms $e_i$, i = 1,2,3. The scheme always culminates in either acceptance or rejection by A and B.

The environment is setup initially, according to the algorithm *Setupalg* ($e_1$) and published to the parties A and B at time $t_1$. During the Commit phase, A uses algorithm *Commitalg* ($e_2$), which encapsulates a message m∈M, along with secret string $S \in_R \{0,1\}^k$ into a string c. The opening key (secret key) could be formed using both m and S. A sends the result c to B( at time $t_2$). In the Open phase, A sends the procedure for revealing the hidden Commitment at time $t_3$, and B uses this.

*Openalg* ($e_3$): B constructs c' using *Commitalg*, message m and opening key, and checks weather the result is same as the commitment c.

Decision making:
If ( c = c' )
Then A is bound to act as in m
Else he is free to not act as m

## 3 - Fuzzy Commitment Formally Defined:

**When would a commitment scheme as in definition 1 become fuzzy?** At the stage of decision making. This result of uncertainties that make crop up during transmission noise. We may formalize the whole process by properly defining it.

Definition 2:

A *Fuzzy Commitment scheme* is a tuple $\{P, E, M, f\}$ Where $M \subseteq \{0,1\}^k$ is a message space which consider as a code, *P* is a set of individuals, generally with three elements A as the committing party, B as the party to which Commitment is made and TC as the trusted party, *f* is error correction function (def. 5) and $E = \{(t_i, e_i)\}$ are called the events occurring at times $t_i$, i = 1,2,3, as per algorithms $e_i$, i = 1,2,3. The scheme always culminates in either acceptance or rejection by A and B.

In the setup phase, the environment is setup initially and public commitment key CK generated, according to the algorithm *Setupalg* ($e_1$) and published to the parties A and B at time $t_1$. During the Commit phase, Alice commits to a message m∈M according to the algorithm *Commitalg* ($e_2$) into string c. In the Open phase, A sends the procedure for revealing the hidden Commitment at time $t_3$ and B use this.

*Openalg* ($e_3$): B constructs c' using *Commitalg*, message t(m) and opening key, and checks weather the result is same as the received commitment t(c), where t is the transmission function.

Fuzzy decision making:
If (nearest(t(c),f(c') )≤$z_0$)
Then A is bound to act as in m
Else he is free to not act as m

## 4-Numerical example:

Let *P* = { Alice, Bob} i.e. we consider a situation where there is not trusted party.

*Message space*: Let M ={0000, 1011, 0101, 1110, 1010, 1100, 1111} $\subset \{0,1\}^4$.

*Message*: let m=1011

*Encoding function*: Let g: M→$\{0,1\}^7$ be one to one function defined as :
g(M) = C ={0000000 = g(0000), 0100101 = g(1011), 0010011=g(0101), 0110110=g(1110), 1011010=g(1010), 1101100 =g(1100), 1111111 = g(1111)} $\subset \{0,1\}^7$

The image set C under g is a code set, which is satisfies the closure property under XOR operation, an element of C is also called a *codeword*.

*Setup phase*: At time $t_1$, it is agreed between all that
CK ≅ XOR
f ≅ nearest neighbour in set C.
$z_0$=0,20.

*Commit phase*: At time 2
Alice committed to her massage m=1011. She knows that g(m)=g(1011)=0100101
For sake of secrecy she selects $S \in_R C$ at random, Suppose S=1011010.
Then her commitment c = *Commitalg*(CK, g(m), S) = g(m) XOR S= 1111111
Alice sends c to Bob, which Bob will receive as t(c), where t is the transmission function. Let the transmitted value t(c) = 1011111, which includes noise.

*Open phase*: At time t3
Alice discloses the procedure g(m) and S to Bob to open the commitment.
Suppose Bob gets t(g(m))= 1100101 and t(s)=1011010.
Bob compute
c'=*Commitalg*(CK,t(g(m)),t(s))=t(g(m))XORt(S)=0111111.
Bob check that dist(t(c),c')=2, he will realize that there is an error occur during the transmission.
Bob apply the error correction function *f* to c':
f(c')=1111111 (the nearest neighbour of c'=0111111 is 1111111).
Then Bob will compute nearness(t(c),f(c'))=dist(t(c),f(c'))/n =1/7 =0.14. (def.6)
Sine 0.14≤$z_0$=0,20.
Then FUZZ(f(c'=0111111))=0 (def.7).
Bob accepted t(c)=f(c')=1111111.
Finally Bob calculate $g^{-1}$(1111111)=1011.

## 5- Error Correcting Codes:

Definition 3: A metric space is a set C with a distance function dist:C×C→$R^+$=[0,∞), which obeys the usual properties (symmetric, triangle inequality, zero distance between equal points).

Definition 4: Let C $\{0,1\}^n$ be a code set which consists of a set of codewords $c_i$ of length n. The distance metric between any two codewords $c_i$ and $c_j$ in C is defined by

$$dist(c_i,c_j)=\sum_{r=1}^{n} | c_{i_r} - c_{j_r} | \quad c_i,c_j \in C.$$

This known as Hamming distance[16]

Definition 5: An error correction function *f* for a code C is defined as
f($c_i$)={$c_j$ | dist($c_i$,$c_j$) is the minimum, over C-{$c_i$}}
Here $c_j$ =f($c_i$) is called the nearest neighbor of $c_i$.

Definition 6: The measurement of nearness between two codewords c and c' is defined by
nearness(c,c')=dist(c,c')/n,
it is obvious that 0≤nearness(c,c') ≤1.

Definition 7: The *fuzzy* membership function for a codeword c' to be equal to a given c is defined as

FUZZ(c')= 0     if nearness(c,c')=z≤$z_0$<1
         =z     other wise

## 6- Real Life Situation :(Testament):

Alice wants to write a testament to declare she passes all her fortune to her son Bob after her death. Of course, the Alice's attorney is playing the role of the authority.

Setup phase: at time $t_1$

     Attorney published to Alice and Bob an envelope as a public commitment key, error correction function *f* and $z_0$

Commit phase: at time $t_2$

     Alice writes her testament m and put it in a sealed envelope (commitment c) and gives to her son Bob. During the time pass some letters of the testament corrupted we assume that it is t(c).

Open phase: at time $t_3$ (death time of Alice)

     Attorney on behalf of Alice meet Bob and reveal to him the original testament (also during the time may be some letters corrupted of the original testament i.e. t(m)), they open the envelope to obtain the testament m', and they calculate

- nearness(t(m),*f*(m'))
- If (FUZZ(m')=0)
       Then m'=m
       Else m'≠m

## 7- Fuzzy Commitment Schemes from Literature:

| Name of the paper | Name of the author | Year of publishing | Concepts used |
|---|---|---|---|
| A fuzzy commitment scheme | A. Juels and Martin W. | Sixth ACM Conference on Computer and Communications Security, pages 28-36, ACM Press. 1999. | Cryptography, Error correcting codes and commitment schemes, fuzzy logic |
| Error-Tolerant password recovery | N.Frykholm and A. Juels | *Eighth ACM Conference on Computer and Communications Security*, pages 1-8. ACM Press. 2001 | Error correcting codes, Cryptography and fuzzy commitment scheme |

## 8- Concluding remarks:

We have attempted to formalize definition of a *fuzzy* commitment scheme by introducing a *fuzzy* membership function at the opening algorithm stage. Introduction of error correction function was introduced by many research workers earlier [16,17,19]. Introduction of the Fuzzy member ship function makes the use of word fuzzy more explicit.

## References

[1] Manuel Blum, Coin flipping by telephone. *Advances in Cryptology: A Report on CRYPTO '81*, pp. 11–15, 1981, http://www.cs.cmu.edu/~mblum/research/pdf/coin/

[2] George J. Klir and Bo Yuan, *Fuzzy Sets and Fuzzy Logic theory and applications*, Prentice Hall of India private limited, New Delhi 2000.

[3] A. Juels and M. Wattenberg. *A fuzzy Commitment Scheme*. In *Proceedings of the 6th ACM Conference on Computer and Communication Security*, pages 28–36, November 1999.

[4] A. Juels and M. Sudan, "*A fuzzy vault scheme*," *Proceedings of IEEE Internation Symposium on Information Theory*, p.408, IEEE Press, Lausanne, Switzerland, 2002..

[5] Torben Pryds Pedersen, Non-Interactive and Information- Theoretic Secure Verifiable Secret Sharing. *Advances in Cryptology - CRYPTO '91, 11th Annual International Cryptology Conference*, pp. 129–140,1991, http://www.cs.cornell.edu/courses/cs754/2001fa/129.PDF.

[6] M.Naor: Bit Commitment using pseudo-randomness, *J. of Cryptology, Volume 4*, pp. 151-158
http://www.wisdom.weizmann.ac.il/~naor/topic.html

[7] Shai Halevi, Silvio Micali, Practical and Provably-Secure Commitment Schemes from Collision-Free Hashing. *Advances in Cryptology - CRYPTO '96, 16th Annual International Cryptology Conference*, pp. 201–215, 1996, http://coblitz.codeen.org:3125/citeseer.ist.psu.edu/cache/papers/cs/778/ftp:zSzzSztheory.lcs.mit.eduzSzpubzSzpeoplezSzshaihzSzcomitmnt2.pdf/halevi96practical.pdf.

[8] Shai Halevi, Efficient Commitment Schemes with Bounded sender and Unbounded Receiver, *Proceedings of Crypto '95 LNCS. Vol.963* Springer-Verlag 1995 pages 84-96. http://citeseer.ist.psu.edu/halevi96efficient.html

[9] Hans Delfs and Helmut Knebl: *Introduction to Cryptography principle and applications* Springer-Verlag Berlin Heidelberg 2002.

[10] William Stallings, 2001: *Network Security Essentials applications and standards,* Wesley Longman (Singapore) Ptd. Ltd. Indian branch.

[11] Alfred Menezes, Paul Van Oorschot and Scott Vanstone: *Handbook of Applied Cryptography* CRC press 1996.

[12] Ivan Damg°ard, Jesper Buus Nielsen, *Commitment Schemes and Zero-Knowledge Protocols*, 2006, http://www.daimi.au.dk/~ivan/ComZK06.pdf.

[13] Eiichiro Fujisaki, Tatsuaki Okamoto, Statistical Zero Knowledge Protocols to Prove Modular Polynomial Relations. *Advances in Cryptology - CRYPTO '97, 17th Annual International Cryptology Conference,* pp. 16–30, 1997, http://dsns.csie.nctu.edu.tw/research/crypto/HTML/PDF/C97/16.PDF

[14] Xavier Boyen. Reusable Cryptographic Fuzzy Extractors. In *11th ACM Conference on Computer and Communications Security (CCS 2004)*, pages 82-91. ACM Press, 2004..

[15] Yevgeniy Dodis, Leonid Reyzin, and Adam Smith, Fuzzy extractors: How to generate strong keys from

biometrics and other noisy data, *In Proceedings of the International Conference on Advances in Cryptology (EUROCRYPT '04), Lecture Notes in Computer Science*. pp. 523-540, Springer Verlag, 2004.

[16] V.Pless, *Introduction to theory of Error Correcting Codes* Wiley, New York 1982.

[17] G.A. Jones and J.M. Jones: *Information and Coding Theory* Springer-Verlag London Limited 2000.

[18] N. Frykholm and A. Juels, Error-Tolerant Password Recovery. In P. Samarati, ed., *Eighth ACM Conference on Computer and Communications Security*, pages 1-8. ACM Press. 2001.

[19] R.J. McEliece, *The Theory of Information and Coding*. Cambridge Univ. Press, 2002.